\documentclass[pra,twocolumn,a4paper,showpacs,floatfix]{revtex4}
\usepackage{graphicx} \newcommand{\figspace}{\vspace*{2ex}}
\usepackage{amsmath}
\begin{document}
\title{Extended electron states and magneto-transport in a 3-simplex fractal}
\author{Arunava Chakrabarti\footnote{On leave from Department of Physics,  
University of Kalyani, Kalyani, West Bengal 741 235, India.}} \affiliation{Max-Planck-Institut f\"{u}r Physik 
Komplexer Systeme \\
N\"{o}thnitzer Strasse 38, D-01187 Dresden, Germany}
\begin{abstract}
We perform a real space renormalization group analysis to study the
one electron eigenstates and the transmission coefficient across a 3-simplex
fractal lattice of arbitrary size, in the presence of a magnetic field. In particular, we
discuss the existence of extended electronic states and the influence of the magnetic
field on the spectrum of the system. A general formulation is presented which enables
us to deal with both the isotropic and the anisotropic cases. It is found that in general, 
the transmission across an anisotropic fractal is larger than its isotropic counterpart.
Additionally, we point out an extremely interesting correlation between a subset of values of the 
external field, and the multiple fixed point cycles of the Hamiltonian corresponding to the 
extended eigenstates. A prescription for observing such correlation is proposed. This aspect
may be important in the classification of the extended states in such deterministic structures. 
\end{abstract}
\maketitle 
\section{Introduction}
The physics of noninteracting electrons on lattices without any translational
invariance has been the subject of intense theoretical activity since many years. 
Beginning with the work of Anderson \cite{pw58} describing the absence of diffusion in 
randomly disordered lattices, a wealth of knowledge has now accumulated in this
field. Over the last couple of decades the problem of electron localization in 
low dimensional quasicrystalline and fractal lattice models have enriched this
field further \cite{jbs85}-\cite{yg81}.

Regular fractal networks have already been appreciated to be linked to 
certain disordered structures, such as the percolation backbone clusters \cite{bm82,yg81}.
Such systems, by the way they are generated, are self similar and exhibit the 
absence of any long ranged translational order. Yet, they are not random. This
gives the fractal networks the status of being intermediate between perfectly
periodic structures and completely disordered ones. Naturally, they became the
objects of detailed theoretical study over all these years.

A regular fractal lattice with finite ramification (meaning that a large part 
can be detached by removing a finite number of bonds) is generally solvable, and 
is found to possess exotic spectral features \cite{edom83}-\cite{sch91}. The properties are similar for
electrons and other excitations such as phonons or magnons. For example, it is 
found \cite{edom83} that, contrary to the absolutely continuous electronic spectrum of a 
perfectly crystalline sample, or the pure point spectrum of a randomly disordered
lattice, the electronic spectrum of a deterministic fractal is typically a 
Cantor set of measure zero. The changing local environment around each lattice 
point leads to localization of the single particle eigenstates, but in a way
not found common in conventional disordered samples. The conductance in such 
fractalline systems is also found to show up power-law scaling with the size of 
the system. The scaling exponent may be distributed in a multifractal way \cite{sch91}, as the 
system approaches its thermodynamic limit.

It is now known, that even a deterministic fractal may support a countable infinity 
of extended eigenstates \cite{xr95}-\cite{bb96}. 
 This observation is interesting, as 
fractal usually does not exhibit any 
short ranged `positional correlation' between the lattice points, as in the cases 
of certain random or quasiperiodic lattice models, known to support extended 
single particle states \cite{dun90}-\cite{ac94}. 
No local clusters of sites can really be identified which lead to resonance causing
an extendedness in the basic character of the eigenfunction.
The delocalization of an  
infinite number of extended single particle states in a fractal can thus be attributed to
the structure of the lattice as a whole.

In this communication we report a real space renormalization group (RSRG) analysis of 
the spectral features of a $3$-simplex fractal \cite{dd77} in the presence of a constant magnetic
field. The growth of a $3$-simplex fractal is described in Fig. 1. We choose the magnetic flux
to penetrate the small triangular plaquettes at each generation, in a 
direction perpendicular to the plane of the fractal. This results in a break in 
the time reversal symmetry as the electron hops along the edges of a basic triangle. The
hopping along the `bond' joining the two neighboring triangle remains unaffected by the 
field.  
The magnetic field is already known to have a non-trivial effect on the Cantor set energy 
spectrum usually supported by such fractals as has been demonstrated in the case 
of a Sierpinski gasket network \cite{jr85}. The degeneracy of the zero field solution is 
found to be broken, and the energy spectrum broadens up, though the isolated character of the 
eigenstates is preserved. Later, Wang \cite{xr96} re-examined the Sierpinski gasket in the
presence of a magnetic field using an RSRG method, but allowed only a subset of the full 
parameter space to evolve under RSRG transformations. The suppression of the evolution of the
full parameter space, to our mind, sometimes may lead to incomplete information about the eigenstates.

We take up the investigation of the spectral properties of a $3$-simplex lattice mainly motivated
by the following ideas: 

First, apart from being representatives of certain percolation clusters, 
deterministic fractal geometries have also been realized experimentally. 
Modern nano-fabrication 
techniques have made it possible to experimentally investigate model systems, such as 
the Josephson junction arrays and superconducting networks developed following a Sierpinski 
gasket fractal \cite{jmg86,sek95} where both the nature, and the amount of disorder can be 
accurately controlled. The level of frustration in such case can be tuned by an external magnetic field.
Additionally, recent experimental measurements of persistent current in an array of mesoscopic rings 
\cite{rab01} suggest that, one can think of a deterministic fractal network with multiple loops at all 
scales of length and observe the interplay of magnetic 
field and the fractal geometry in the transport and related issues in such networks. 
Detailed study in this regard  
has received relatively little attention so far \cite {ac97}, and deserves more analysis.

Second, we wish to have a deeper look at whether the external magnetic field, if it generates
extended eigenfunctions for the fractal, is linked to any cyclic invariance 
of the  
Hamiltonian. If such cycles at all exist, it is important to
understand whether they have any correlation with the values of the
external flux responsible for them. 
In that case it would be possible to tune the flux, which is an external {\it agent} to 
control the transport in finite fractal networks, as well as a classification between 
different extended states could be achieved. This aspect which has 
remained, to the best of our knowledge, really un-explored so far, is a major focus of the present
work.

We find several interesting features. In the absence of any field, and with the introduction 
of anisotropy in the 
amplitude of electron hopping, 
the corner - to - corner transmission of a
$3$-simplex fractal turns out to be better than its isotropic counterpart. Clusters of 
high transmittivity pack the transmission spectrum in the former case, whereas, the 
system turns out to be poorly transmitting in the isotropic situation. 
This remains the general feature for both the isotropic and the anisotropic cases 
when a magnetic field is turned on, with the transmission coefficient exhibiting 
interesting Aharonov-Bohm (AB) oscillations. 

The magnetic field is also found to have a dramatic effect 
on the local density of states at a corner site of an infinite (or, semi-infinite)
$3$-simplex fractal. An `apparently' continuous distribution of eigenvalues appear around the 
centre of the spectrum, suggesting the formation of a `band' of extended states. 
We have not been able to
prove conclusively the existence of a `band' of extended states. However, extensive numerical
investigation of the local density of states and the behavior of the nearest neighbor hopping
integrals under renormalization group iterations are suggestive of the fact.

Finally, we have particularly focused on a specific value of the energy of the electron
which supports an extended eigenstate only in the presence of a magnetic flux, 
and have carried out careful numerical investigations on the flow of the parameters of the 
Hamiltonian under repeated renormalization of the system, as the 
flux is changed systematically. We find that, for this specific
extended eigenstate, values of the magnetic flux show up a strong correlation with the 
fixed point cycles of the parameter space, that can be predicted successfully. Such observations 
 may provide an idea of classifying the extended states in regards of the fixed points of the
Hamiltonian  brought about by the external magnetic field.

In what follows, we report our results. In section II, we introduce the model and the RSRG 
scheme. Section III contains a discussion on the nature of the energy spectrum both in the 
absence and the presence of a field.
Section IV is devoted to the investigation of
magneto-transport, while in section V we discuss how the flux values 
can be correlated to the cycles of the fixed point. In section VI we draw our conclusions.
\vskip .3in
\section{The model and the RSRG scheme}
We begin by referring to Fig. 1. The three basic triangular plaquettes (Fig. 1(a)) are 
placed as shown in the figure  to generate a second-generation fractal (Fig. 1(b)). The process
continues.
A magnetic field penetrates each small triangle. 
We work within a tight binding formalism in which the Hamiltonian for the 
electron in a basic triangular plaquette is written as, 
\begin{equation}
H = \sum_{i} \epsilon_i c_i^{\dag} c_i + \sum_{<ij>} t_{ij} e^{i\theta} c_i^{\dag} c_j
\end{equation}
In the above, $\epsilon_i$ is the on-site potential which, in the most general 
anisotropic model, can assume two values, $\epsilon_\alpha$ (at the corner sites
at the horizontal base of each elementary triangle) and $\epsilon_\mu$ at the remaining vertices. 
The nearest neighbor hopping integrals $t_{ij}$ are assigned amplitudes $t_x$ and $t_y$ for hopping
across the horizontal and the angular bonds within each elementary triangle respectively. 
The inter-triangle connection is given by $t_{ij}=T_x$ in the horizontal direction, and 
$t_{ij}=T_y$ otherwise, as depicted in Fig. 1(b), and these are 
free from any associated flux.
It is to be appreciated that the status of a vertex ($\alpha$ or $\mu$) is governed only by
the bonds $t_x$ or $t_y$ attached to it. $T_x$ and $T_y$ remain un-decimated on renormalization, 
and does not play a part in fixing the status of a vertex.
It may also be noted at this point that,  
this  $3$-simplex network differs, in the 
presence of a magnetic field penetrating its elementary plaquettes, from its closest look-alike
the Sierpinski gasket \cite{xr96} in the fact that in the $3$-simplex network, the time reversal symmetry of the 
electron hopping is broken only partially, i.e., when the electron hops along the edges of an 
elementary triangle, compared to a Sierpinski gasket, where it is broken uniformly. 
It will be interesting to see the consequence of this at various scales 
of length.

The magnetic flux threading each small
triangle enters the Hamiltonian only through the hopping integrals along the sides of the
elementary triangles \cite{jr85}. We set $\theta=2\pi \Phi/(3\Phi_0)$. Here, $\Phi$ is the flux threading
each small triangle, and $\Phi_0$ is the flux quantum.
Following Banavar, Kadanoff and Pruisken \cite{jr85}, we
select the gauge in such a way that the a factor of $\exp(i\theta)$ is associated with
either $t_x$ or $t_y$ when the electron hops in the direction given by the arrow. The 
phase is opposite when it hops back. 
Thus, the `forward' or the `backward' hopping integrals in the presence of the field will be 
written as, $t^{F(B)}_{x(y)} = t_{x(y)} e^{\pm i\theta}$.
The model, as it is presented, allows one to study
a perfectly general anisotropic case ($t_x \ne t_y$), 
including a hierarchical distribution of the bonds $T_x$ or $T_y$, 
which is known to exhibit a very interesting `restoration of isotropy' in a 
$3$-simplex network of classical resistors \cite{lin96}, although, we 
do not pursue this topic here.  The isotropic limit is easily retrieved as
well.
\begin{center}
\begin{figure}
\centering \figspace
\centerline{\includegraphics{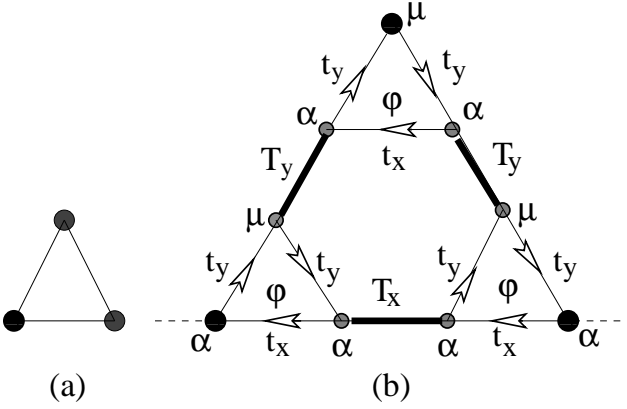}}
\caption{\label{fig1}(a) The basic building block of a $3$-simplex network. (b) The lattice in its
second generation. The dotted lines indicate the leads attached for transmission studies. }
\end{figure}
\end{center}
The sites marked with decorated circles 
are now decimated to yield the following recursion
relations for the on-site potentials and the hopping integrals.
\begin{eqnarray}
\epsilon_{\alpha,n+1} & = & \epsilon_{\alpha,n} + C_{1,n} t_{y,n}^F + D_{1,n} t_{x,n}^B \nonumber \\
\epsilon_{\mu,n+1} & = & \epsilon_{\mu,n} + C_{4,n} t_{y,n}^B + D_{4,n} t_{y,n}^B \nonumber \\
t_{x,n+1}^F & = & C_{3,n}^* t_{y,n}^B  + D_{3,n}^* t_{x,n}^F \nonumber \\
t_{y,n+1}^F & = & C_{2,n} t_{y,n}^F + D_{2,n} t_{x,n}^B \nonumber \\
t_{x,n+1}^B & = & {t_{x,n+1}^F}^* \nonumber \\
t_{y,n+1}^B & = & {t_{y,n+1}^F}^*
\end{eqnarray}
In the above, $n$ and $n+1$ in the subscript denote the stages of renormalization 
, ${t^{F(B)}}_{j,n} = t_{j,n} e^{\pm i\theta_n}$ with $j=x$ or $y$ 
representing the $x$- or $y$-hopping along the edge of a triangle, and,  
$\theta_n$ is the `re-normalized' flux at the $n$th stage, which of course, we 
do not have to calculate separately.
Here,  
\begin{eqnarray}
C_{1,n} & = & \frac{A_{1,n}+A_{2,n}B_{1,n}}{1 - A_{2,n} B_{2,n}} \nonumber \\
C_{2,n} & = & \frac{A_{3,n}+A_{2,n}B_{3,n}}{1 - A_{2,n} B_{2,n}} \nonumber \\
C_{3,n} & = & \frac{A_{4,n}+A_{2,n}B_{4,n}}{1- A_{2,n} B_{2,n}} \nonumber \\
C_{4,n} & = & \frac{B_{5,n}+A_{5,n}B_{6,n}}{1 - A_{6,n}B_{6,n}} \nonumber \\
D_{1,n} & = & \frac{B_{1,n}+B_{2,n}A_{1,n}}{1 - A_{2,n}B_{2,n}} \nonumber \\
D_{2,n} & = & \frac{B_{3,n}+B_{2,n}A_{3,n}}{1 - A_{2,n}B_{2,n}} \nonumber \\
D_{3,n} & = & \frac{B_{4,n}+B_{2,n}A_{4,n}}{1 - A_{2,n}B_{2,n}} \nonumber \\
D_{4,n} & = & \frac{A_{5,n}+B_{5,n}A_{6,n}}{1 - A_{6,n}B_{6,n}} 
\end{eqnarray}
At any stage $n$ of renormalization, 
we have defined (suppressing the subscript $n$),  
\vskip .22in
$A_1=r t_y^B/s$, $A_2=(r t_y^F + T_xT_y^2t_x^Bt_y^B)/s$, 
$A_3=T_y(qt_y^F + pt_y^Bt_x^B)/s$, $A_4=T_y^2t_x^B [(E-\epsilon_\alpha) t_y^F + t_x^Bt_y^B]/s$,
$A_5=t_y^Bz_1/[z_1(E-\epsilon_\alpha) - T_y^2]$,
and, $A_6=(z_1t_x^F + v_1T_yt_y^B)/[z_1(E-\epsilon_\alpha) - T_y^2]$.
\vskip .22in
$B_1=t_x^F/z$, $B_2=(rt_y^B + T_xT_y^2t_x^Ft_y^F)/(rz)$, 
$B_3=uT_xT_yt_y^F/(pz)$, $B_4=T_x[wT_yt_y^F + t_x^B(E-\epsilon_\mu) + (t_y^F)^2]/(pz)$,
$B_5=t_y^F/z_2$, and, $B_{6}=(T_yw_2+t_x^B)/z_2$. 
\vskip .25in

The quantities $u$, $w$, $z$, $v_1$, $z_1$, 
 $w_2$ and $z_2$  are defined as, 
\begin{eqnarray}
u & = & \frac{p \left [t_x^F(qt_y^F + pt_y^Bt_x^B) + rt_y^B \right ]}{qr} \nonumber \\
w & = & \frac{T_y [t_x^Bt_y^B + (E-\epsilon_\alpha)t_y^F]}{qr}  (p t_x^Ft_x^B+r) \nonumber \\ 
z & = & (E-\epsilon_\alpha) - \frac{T_x[(E-\epsilon_\mu)T_x + t_y^FT_yv]}{p} 
\end{eqnarray}
with, 
$v=[pT_xT_yt_x^Ft_x^Bt_y^B + rT_xT_yt_y^B]/(qr)$,
\begin{eqnarray}
v_1 & = & \frac{T_xT_yt_y^B}{p(E-\epsilon_\alpha)-T_x^2(E-\epsilon_\mu)} \nonumber \\
z_1 & = & (E-\epsilon_\mu) - w_1t_y^B \nonumber \\
\end{eqnarray}
with, $w_1 = t_y^Fp/[p(E-\epsilon_\alpha)-T_x^2(E-\epsilon_\mu)]$,
and, 

\begin{eqnarray}
w_2 & = & \frac{(t_y^F)^2 T_xT_y}{(E-\epsilon_\mu)[p(E-\epsilon_\alpha)-T_x^2(E-\epsilon_\mu)]
-pt_y^Ft_y^B} \nonumber \\
z_2 & = & (E-\epsilon_\alpha) - T_yv_2
\end{eqnarray}
with,
$v_2=\frac{T_y[p(E-\epsilon_\alpha)-T_x^2(E-\epsilon_\mu)]}
{(E-\epsilon_\mu)[p(E-\epsilon_\alpha)-T_x^2(E-\epsilon_\mu)]
-pt_y^Ft_y^B}$.
\vskip .23in
Finally, the remaining factors are, 
\begin{eqnarray}
p & = & (E-\epsilon_\alpha) (E-\epsilon_\mu) - t_y^Ft_y^B \nonumber \\
q & = & (p-T_y^2) (E-\epsilon_\alpha) \nonumber \\
r & = & q(E-\epsilon_\alpha) - pt_x^Ft_x^B \nonumber \\
s & = & r(E-\epsilon_\mu) - qT_y^2 \nonumber \\
\end{eqnarray}

The inter-triangle hopping integrals $T_x$ and $T_y$ remain un-affected as a 
result of renormalization, i.e,. $T_{j,n+1} = T_{j,n}$ at any stage $n$, 
$j$ representing $x$ or $y$.

The set of recursion relations given by Eq. (2) is a highly non-linear one, and 
can be reduced to a simple form only under simple isotropic model in the absence of
any field. However, they are not so difficult to  deal with numerically,  
and yield quite
a few interesting results which we shall now discuss.
\vskip .3in
\section{Energy spectrum and the nature of eigenstates}
\subsection{The zero field case}
To begin with, we have evaluated the local density of states (LDOS) at a corner site of
a $3$-simplex gasket in the isotropic limit, and in the absence of any magnetic field. We
set $\epsilon_\alpha = \epsilon_\mu = \epsilon$, $t_x = t_y = t$, and, $T_x = T_y = \tau$. 
The recursion relations given by Eq. (2) now get reduced to, 
\begin{eqnarray}
\epsilon_{n+1} & = & \epsilon_n + \frac{P_n}{Q_n R_n} \nonumber \\
t_{n+1} & = & \frac{U_n}{Q_n R_n}
\end{eqnarray}
with, 
$P_n = 2{t_{n}}^2[(E-\epsilon_n)^2-\tau (E-\epsilon_n)-t_n^2]$, 
$Q_n = (\tau+t_n+\epsilon_n-E)$, $R_n = \tau^2-(E-\epsilon_n)^2+t_n^2-\tau t_n$ and, 
$U_n = \tau t_n^2(E-\epsilon_n+t_n-\tau)$. $n$ again stands for the stage of renormalization.
In the absence of any field, it is a simple task to
evaluate the LDOS at a corner site using the standard Green's function technique \cite{bw83}. 
We have checked that the recursion relations produce the correct LDOS at the corner site of 
a one dimensional chain \cite{ad97} in the limit $t_y, T_y \rightarrow 0$.
With a 
small imaginary part added to the energy $E$, the hopping integral $t$ flows to zero after 
certain steps of RSRG. 
This implies that the lattice, 
at that scale of length, breaks up into an assembly of diatomic molecules, with the
hopping $\tau$ connecting the `atoms' remaining unchanged. 
Each such molecule will be decoupled from its neighbors, and the transmission across 
the lattice will be zero.
The on-site potential at the corner site reaches its
 fixed point value $\tilde\epsilon$, and, the LDOS (meaningful only at an 
extreme corner site, which is truly decoupled from the rest of the lattice) is then evaluated as, 
\begin{equation}
\rho(E) = -\frac{1}{\pi} G_{00}(E+i\eta) 
\end{equation}
where, the diagonal Green's function $G_{00} = 1/(E-\tilde\epsilon)$. A plot of the LDOS 
at a corner site is shown in Fig. 2 (top figure). The LDOS is exhibited is checked, as permitted by 
the limit of accuracy,  to 
be stable against a decreasing value of the imaginary part of the energy $E$, 
and has been displayed within a value unity to
give prominence to the smaller peaks compared to the much bigger ones. The fragmented, scanty
appearance is consistent with the usual Cantor set spectrum common to such systems. 

Before introducing
the magnetic field, it is pertinent to comment on the existence of extended eigenstates in this
simplified model. 
If the recursion relations Eq. (8) are iterated for any arbitrary energy with no imaginary part 
added to it, 
the hopping integral $t$ flows to zero after  certain steps of RSRG. 
This means that the corresponding energy should lie in a gap
of the spectrum of the infinite system, or, corresponds to an exponentially localized
eigenstate, though it is not possible to distinguish between these two cases by simply
looking at the flow of the hopping integral. 
It should be remembered that a zero LDOS is not a conclusive proof for an energy not to be 
in the spectrum of the infinite lattice. 
On the other hand, if, for certain energy, the
hopping integral remains non-zero under an indefinite number of iterations, then we have
definitely hit upon an extended eigenstate. It is not difficult, using Eq.s (8) to fix an
energy for which $\epsilon_{1}=\epsilon$. The energy, evaluated in this fashion can then be 
inserted into the second equation in (8), and one can select $T$ in such a manner that $t_1$
under RSRG remains equal to $t$. As the inter-triangle hopping is always $T$, we thus have a 
fixed point of the parameter space, viz, $(\epsilon_n,t_n,T_n) \rightarrow (\epsilon_{n-1},t_{n-1},T_{n-1})$
for $n \ge 1$. 
The corresponding state will be extended in nature. This of course demands a definite
relationship between $\epsilon$, $t$ and $T$ , that is, we are 
talking of a specific model. For example, with $\epsilon=0$, 
and $t=1$, if we set $E=0.5t$, then the fixed point behavior sets in from the first RSRG
step onwards if $T$ is assigned a value equal to $-3/2$ (in units of $t$). Much more 
involved relationship between the parameters of the system is capable of leading to a 
fixed point behavior starting at deeper scales of renormalization. However, in each case, 
a different set of parameters essentially means that we are dealing with a different system.
In principle, we will observe extended eigenstates in each of those cases which provide a 
meaningful relationship between the parameters, but only at certain discrete energy values.
\vskip .25in
\begin{center}
 
\begin{figure}
\centering \figspace
\centerline{\includegraphics[width=0.95\columnwidth,angle=-90]{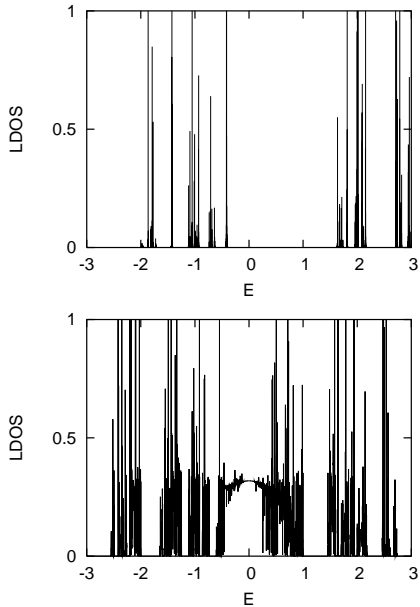}}
\caption{\label{fig2}The local density of states at a corner site of an isotropic $3$-simplex fractal. The top figure
corresponds to the case of zero flux, whereas the bottom one corresponds to the case where 
the flux $\Phi=\Phi_0/4$. 
We select $\epsilon=0$, and, $t=1$.
The imaginary part added to the energy is $10^{-5}$ in unit of $t$.}
\end{figure}
\end{center}
\subsection{A magnetic field is turned `on'}
A magnetic field, however, is capable of bringing dramatic changes into the spectrum
even when we deal with a particular system with a pre-defined set of parameters.
In the lower one in Fig. 2 the LDOS at the corner site is shown for a flux $\Phi=\Phi_0/4$ 
for the simplest isotropic model with $\epsilon_\alpha=\epsilon_\mu=0$, and 
$t_x=t_y=T_x=T_y=1$. The spectrum
shows very closely spaced zones of finite density of states. Of particular interest is the
narrow energy interval around the `centre' $E=0$, where, the appearance of a smooth region 
of almost constant density of states exits. It has not been possible to prove the existence
of a continuous {\it band} of states, though we have made very fine scan around $E=0$,
each time reducing the energy interval to be scanned and diminishing the imaginary part to be 
added to the energy $E$. The LDOS is found to be stable under a variation in the imaginary part $\eta$
from $10^{-3}$ to $10^{-10}$ (in unit of $t$), and within the limits of machine accuracy, it is tempting to
conjecture the existence of a continuous zone of eigenstates around $E=0$. The occurrence of 
dense clusters of non-zero density of states has been tested with other values of the magnetic 
field. The general conclusion is the same (except for $\Phi=m\Phi_0/2$, $m$ being an integer) 
, and the qualitative features do not essentially change 
even when we deal with the anisotropic gasket.
\section{Transmission across finite 3-simplex fractals}
To calculate the quantum mechanical transmission across a finite $3$-simplex 
network of any size, we adopt the very well known formalism proposed by Douglas Stone et al . 
\cite{stone81}, and consequently used by others as well \cite{bb96}. The essential method consists 
in placing the desired fractal network between perfectly ordered semi-infinite leads 
(shown by dashed lines in Fig.1(b)) connected 
to the two extreme $\alpha$-sites at the base. The leads may be described by a uniform 
on-site potential $\epsilon_0$, and constant nearest neighbor hopping integral $t_0$. 
A network at the $n$th generation is then renormalized $n-1$ times to reduce it to
a simple triangle, and finally to a diatomic `molecule' \cite{bb96}, 
still clamped between the leads,  with an effective
on-site potential $\epsilon_{eff}$, 
and an effective hopping integral $t_{eff}^{F(B)}$.
 The transmission coefficient is then 
easily obtained in terms of the quantities $\epsilon_{eff}$, $t_{eff}^{F(B)}$, 
$\epsilon_0$, $t_0$ and the electron energy $E$. The method is so well known that we skip
the detailed mathematical expressions to save space, and present the results only. 
\subsection{Zero flux situation}
Let us start with the isotropic case, that is, $\epsilon_\alpha=\epsilon_\mu$, and, 
 $t_x=t_y=T_x=T_y$. Consider no flux, i.e., $\Phi=0$. The transmission spectrum 
consists of the expected isolated peaks, consistent with the LDOS spectrum. With increasing size the 
fractal turns out to be even poorly transmitting. Interesting changes however start
showing up with the introduction of anisotropy. 
Quite arbitrarily, we set , $t_x=T_x=1$ and 
start reducing the values of
$t_y(=T_y)$ from one towards zero (the limit when the fractal reduces to a 
linear chain clamped between the leads). 
The on-site potentials $\epsilon_\alpha$ and $\epsilon_\mu$ are chosen to be equal, and have 
been set equal to zero.
With the introduction of anisotropy,
regions of finite transmission increase in number. For example, putting $t_y=T_y=0.9$
(that is, a small departure from isotropy), the spectrum is still very much like the
isotropic situation, with new clusters of appreciable transmittance (sometimes unity
as well) appearing in many places. With gradual decrease in the values of the $y$-hopping, 
the newly generated small spiky zones increase in number, join `hand in hand' and the
shape of the entire spectrum start drifting towards what it should be in the case of a 
periodic chain clamped between the leads. The spectrum tends to be restricted within
$E=\epsilon_0 \pm 2t_0$, which is the allowed band of the ordered lead, 
and resembles the spectrum of a $1$-d chain, as $t_y=T_y$ becomes vanishingly small. 
In Fig. 3 we show the transmission spectrum of a $7$-th generation $3$-simplex structure, 
both for the isotropic situation (top), and the anisotropic cases (the middle and 
bottom ones) in
support of the above remarks.
\begin{figure}
\centering \figspace
\centerline{\includegraphics[width=0.95\columnwidth,angle=-90]{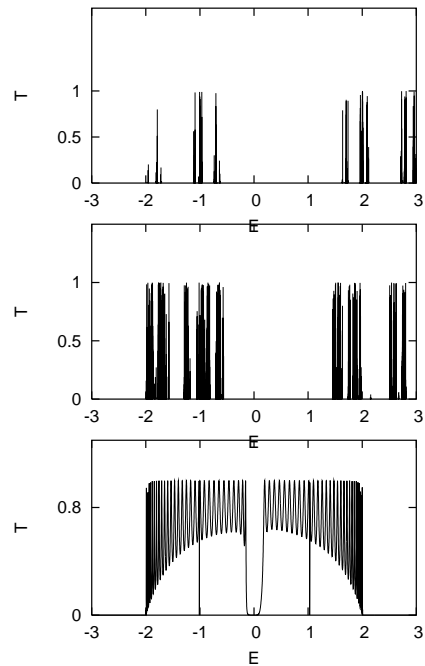}}
\caption{\label{fig3}The transmittivity ($T$) 
against energy of a seventh generation $3$-simplex fractal. The top figure
corresponds to the isotropic case with $t_x=t_y=T_x=T_y=1$.
The middle and the bottom ones depict the situation with $t_y=T_y=0.9$ and $t_y=T_y=0.1$
respectively. The on-site potentials are kept constant at $\epsilon_\alpha=\epsilon_\mu$
in each case. The lead parameters are, $\epsilon_0=0$, and $t_0=2$.} 
\end{figure}
\subsection{Influence of the magnetic field}
We have investigated the general features of the transmission coefficient
as the flux through each elementary plaquette is varied. 
The features are of course
sensitive to the energy of the electron that enters the system through the lead. 
For calculation, we have chosen $\epsilon_\alpha=\epsilon_\mu=0$, and 
$t_x=T_x=t_y=T_y=1$. The lead parameters in this case are chosen to be 
$\epsilon_0=0$, and $t_0=2$ to encompass the full spectrum of the fractal network.
In the case of an isotropic $3$-simplex network, the magnetic field is 
found to generate clusters of resonant transmission throughout the spectrum, a 
particularly noticeable broadening taking place at and around $E=0$. 
\begin{figure}
\centering \figspace
\centerline{\includegraphics[width=0.95\columnwidth,angle=-90]{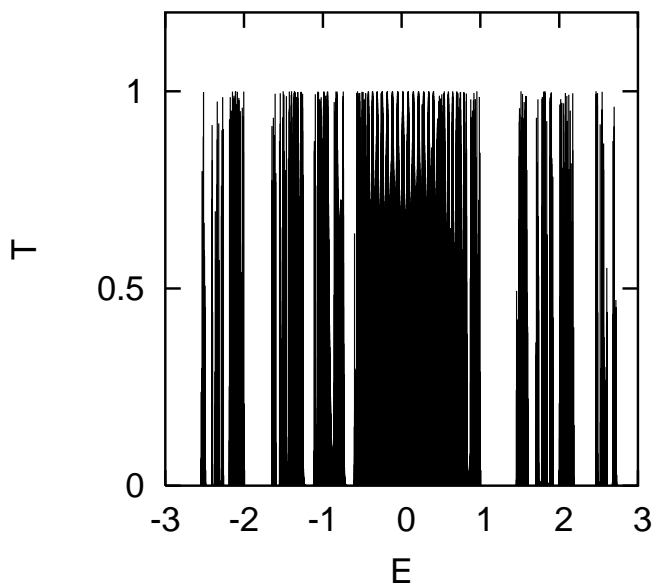}}
\caption{\label{fig4}The transmittivity ($T$) against energy of a sixth generation $3$-simplex fractal. The figure
corresponds to the isotropic case with $t_x=t_y=T_x=T_y=1$.
The on-site potentials are kept constant at $\epsilon_\alpha=\epsilon_\mu=0$, and the
magnetic flux threading each elementary triangular plaquette is $\Phi=\Phi_0/4$. For the lead, 
we take $\epsilon_0=0$, and $t_0=2$.}
\end{figure}
In Fig. 4 we display the transmission spectrum of a sixth generation fractal with energy $E$
with a flux $\Phi=0.25\Phi_0$ threading each elementary triangle. The spectrum is notable
for a thick population of high transmission values at and around $E=0$. Several other {\it mini bands} of
resonant transmission also mark the spectrum. This feature is in sharp contrast to that in the
absence of any field (Fig. 3, the top figure for example). We have observed the change in the 
width of the cluster of high transmission zone around $E=0$ by carefully scanning the energy
interval. It is seen that the width of the transmission window around $E=0$, for any generation, 
changes continuously with the introduction of the flux ,
from zero at $\Phi=0$ to a maximum at $\Phi=\frac{\Phi_0}{4}$, and 
then shrinks back to zero again at $\Phi=\frac{\Phi_0}{2}$. The change in width of this central transmitting
window shows a periodic variation with flux. The period is equal to a half flux quantum. 
Similar observations as above are made with anisotropy in the hopping integrals, but with 
no real new qualitative features. The portion of the $T$ - $\Phi/\Phi_0$ graph immediately 
around $E=0$ remains densely packed with increasing size of the network.

The second half of the study of transmission spectrum consists of an examination of 
the Aharonov-Bohm (AB) oscillations in the transmission coefficient at a fixed
energy of the electron. We have
displayed results for $E=0$. The period of oscillations is found to be equal to one flux
quantum. The detailed features of the spectrum are of course sensitive to the chosen energy 
of the electron. In Fig. 5 we show the AB oscillations 
within one period 
for a third and a sixth generation 
gasket. Once again, the transmission window between zero and one-half flux quantum shows
multiple resonance peaks in the sixth generation fractal compared to a fairly broad and 
structureless shape observed in the third generation. Interestingly, with increasing generation, 
the spectrum is enriched by the appearance of multiple peaks with transmission equal to one (or, 
very close to one), but, the transmittivity really doesn't fall to zero for any appreciable value of the flux
between $0 < \Phi < \Phi_0/2$, and, $\Phi_0/2 < \Phi < \Phi_0$. 

The sensitivity of the spectral features on the parameters of the system are easily revealed 
when we look at the variation of the transmission coefficient against changing magnetic flux 
in the case of an anisotropic simplex lattice. While the electron with $E=0$ doesn't 
distinguish between an isotropic and an anisotropic fractal, other energy values may lead to 
gross changes in the fine structure of the spectrum. 
We point out  an interesting phenomenon.
It is possible to fix up the electron energy in such a way, that, with large anisotropy (that is, 
for low enough values of $t_y=T_y$) , the amplitudes of the AB oscillations start decreasing as the 
transmission coefficient assumes values very close to one. At one stage, the transmittance $T$ becomes 
practically indistinguishable from unity, as $t_y=T_y$ is brought below certain value. By magnifying the 
scale of observation, it is still possible to see how the system tries to preserve the AB oscillations, 
which soon smoothes out if the $Y$-hopping is diminished further. 
In such cases, with increasing generation, the hopping integral $t_{y,n}$ flows to zero, while $t_{x,n}$
does not. It means, as the system grows in size, the intricate geometry of the fractal starts `disappearing' 
to the incoming electron. It essentially feels an ordered chain, and if the energy chosen happens to 
lie in the allowed band of that ordered chain, we get a ballistic transport. 
In Fig. 6 we display one such example where, $E=0.5$ in unit of $t_x$.
\begin{figure}
\centering \figspace
\centerline{\includegraphics[angle=-90]{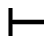}}
\caption{\label{fig5}The transmittivity ($T$) against flux of a third generation (dotted line) 
and a sixth generation (solid line) $3$-simplex fractal. The figure
corresponds to the isotropic case with $t_x=t_y=T_x=T_y=1$ and $\epsilon_\alpha=\epsilon_\mu=0$.
We have set $E=0$.}
\end{figure}
\section{Field induced extended states}
\subsection{General remarks}
The existence of extended eigenstates in systems without any translational
order has always been an intriguing feature in the study of disordered systems. 
However, the complex forms of the recursion relations make an analytical attempt 
rather difficult for a $3$-simplex network. We have therefore relied on a careful numerical study.  
We specially observe the flow of the hopping integrals
under successive RSRG steps. The non-zero values of the hopping integrals under RSRG stand out to
be a definite signature of the state being extended. In all the discussion that follows, we confine
ourselves to the isotropic case only. 
\begin{figure}
\centering \figspace
\centerline{\includegraphics[angle=-90]{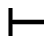}}
\caption{\label{fig6}The transmittivity ($T$) against flux of a $7$-th generation fractal. 
The figure
corresponds to the anisotropic cases with $t_x=T_x=1$, $\epsilon_\alpha=\epsilon_\mu=0$, 
and, $t_y=T_y=0.25$ (solid line), $t_y=T_y=0.20$ (dashed line), and, $t_y=T_y=0.15$ (dotted line).
We have set $E=0.5$ in unit of $t_x$.}
\end{figure}

It is now evident that, the  
magnetic field generates a dense set of eigenvalues, almost resembling a continuum, 
around $E=0$. Quite arbitrarily we have selected a portion $-0.1 \le E \le 0.1$, 
in unit of $t_x$ in a model with $\epsilon_\alpha=\epsilon_\mu=0$, 
$t_x=T_x=t_y=T_y=1$, and have chosen  
the value of the magnetic flux $\Phi=\Phi_0/4$. The LDOS in 
this portion is found to be very stable as the imaginary part added to the energy is decreased
 from $10^{-3}$ to $10^{-9}$. A fine scan over the points in this interval reveals that, any 
energy we hit upon quite randomly in this range, 
 corresponds to a non-zero value of the nearest neighbor hopping 
integral under successive renormalization. This indicates the presence of extended eigenstates. 
Of course, this is not a conclusive proof of the existence of a {\it band} of extended states, 
but, the smaller and smaller widths of the energy interval chosen, remaining close to $E=0$,  
lead to similar behavior. This is 
suggestive of the fact that the presence of a band of extended eigenstates may not be a remote
possibility in this case. We have carried out the numerical investigation for other values of the 
flux and other energy intervals, and in many occasions similar observation has been made. The
observations are in accord with the transmission spectrum displayed in Fig. 5.
\subsection{Values of Flux and cycles of the fixed point: an interesting correlation}

Let us now turn to a different aspect of the problem. 
We restrict the discussion to the isotropic model with $\epsilon_\alpha=\epsilon_\mu=\epsilon$, and 
$t_x=T_x=t_y=T_y=t$. The phase associated with the hopping, at any $n$th stage of RSRG will be 
denoted by $\theta_n$.
We draw the attention of the reader to Fig. 7 which presents the transmission as a function of 
the applied flux for a seventh generation $3$-simplex fractal (solid line), 
together with the LDOS (dashed line) at $E=0$, 
plotted by varying the flux from zero to a single flux quantum. It is interesting to see how the
 energy $E=0$ is periodically brought in and out of the spectrum of the infinite system 
by the magnetic field, the
period being equal to $\Phi_0/2$. Once again, we have tested the robustness of the LDOS spectrum
by diminishing the imaginary part added to energy from bigger $(10^{-3})$ to much smaller $(10^{-9})$
values, so that it is not unjustified to conclude that we definitely have a state at $E=0$. The spectrum
also gives us an indication that we have a continuous LDOS as flux changes from zero to half flux quantum.
We fix our energy of interest at $E=0$.
Looking at the evolution of the hopping integral under successive RSRG steps as we gradually 
increase the flux threading an elementary triangular plaquette from zero to $\Phi_0/2$ and 
beyond, we make two interesting observations. First, for $0 < \Phi < \Phi_0/2$, and, 
$\Phi_0/2 < \Phi < \Phi_0$, the hopping integrals do not flow to zero under iteration, bringing
out the fact that $E=0$ corresponds to a perfectly extended eigenstate for what appears to be a 
continuous distribution of flux values. 

Second, different flux values chosen for the above observation 
unravel the existence of fixed points with multiple cycles. Most interestingly, we have found that the
values of the magnetic flux leading to these cyclic fixed points group into a definite pattern, the 
formation of which can be predicted from the results of our numerical scan of the range of the flux values.
The same pattern (of the number of cycles) is repeated as we change the flux
at certain specially chosen equal intervals along the flux line from $\Phi=0$ to $\Phi=\Phi_0/2$.
We clarify the meaning of the above statements below by citing specific results.

\begin{figure}
\centering \figspace
\centerline{\includegraphics[angle=-90]{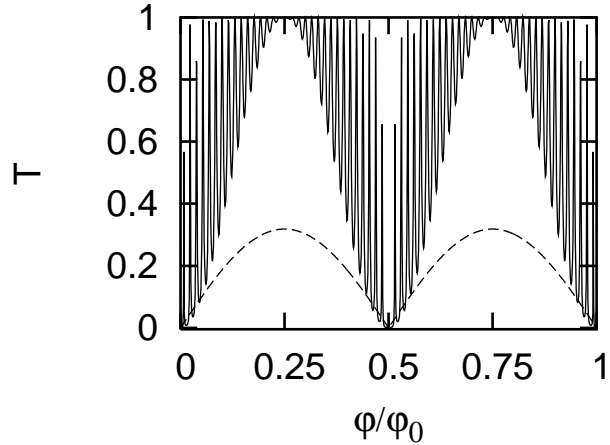}}
\caption{\label{fig7}The transmittivity ($T$) against flux of a seventh generation fractal (solid line) 
and the local density of states at a corner site of an infinite lattice (dashed line) at $E=0$.
We address the isotropic case with $t_x=t_y=T_x=T_y=1$ and $\epsilon_\alpha=\epsilon_\mu=0$.}
\end{figure}

Let us consider the interval $\Phi=0$ to $\Phi=\Phi_0/2$, and divide this interval into 
$2^m$ subintervals with $m$ being a positive integer, and $m \ge 2$. If we exclude the 
two extreme points at $\Phi=0$ and $\Phi=\Phi_0/2$ on this `flux line', then the remaining
values of the magnetic flux making this division are distributed at the locations (coordinates) 
$\Phi_j=\frac{j}{2^{m+1}}\Phi_0$, $j=1,2,3,....,2^m-1$. Let us pick up the simplest choice, viz,  $m=2$. 
The interval $0 < \Phi < \Phi_0/2$ is 
then split into four equal sub-intervals. The values of the flux inside the line (that is, excluding the
values at the boundary) are located at $\Phi_1=\Phi_0/8$, $\Phi_2=\Phi_0/4$ and $\Phi_3=3\Phi_0/8$, 
sequentially from the left. We set $\Phi=\Phi_1=\Phi_0/8$. It is immediately observed that, 
the entire parameter space defined by the trio $(\epsilon_n, t_n, \theta_n)$ gets locked into a 
$2$-cycle fixed point beginning at $n=3$, where $n$ represents the RSRG step. That is, we have
for $\Phi=\Phi_0/8$, 
\begin{displaymath}
 ( \epsilon_n, t_n, \theta_n ) =  ( \epsilon_{n+2}, t_{n+2}, \theta_{n+2} ), n \ge 3
\end{displaymath}
With $\Phi=\Phi_2=\Phi_0/4$, we again get a $2$-cycle fixed point of the parameter space 
$(\epsilon_n, t_n, \theta_n)$. But now, the fixed point behavior is observed for $n \ge 2$.
Setting the external flux equal to the remaining value in this interval, viz, $\Phi=\Phi_3=3\Phi_0/8$, 
we get a $1$-cycle fixed point of the same parameter space beginning at $n=1$, that is, at the 
first stage of RSRG onwards. It should be appreciated that, the values of the flux $\Phi=\Phi_0/8$ and 
$\Phi=\Phi_0/4$ can be termed `equivalent' only in the {\it number of cycles of the fixed point} they
generate. The extended wavefunctions they represent, are characteristically 
different as the invariant cycles of the parameter space set in at different stages of renormalization.

We now increase the value of $m$ in steps. The behavior of the parameter space defined by 
$(\epsilon_n, t_n, \theta_n)$ is observed under successive RSRG iteration as the magnetic flux is 
made to assume sequential values given by $\Phi=\frac{j}{2^{m+1}}\Phi_0$ from the left. Our `experiment'  
leads  to the following very interesting observations:

(i) If we denote the flux leading to
 an $n$-cycle fixed point by $\Phi[n]$, then the values of the flux between zero and the half 
flux quantum 
group themselves into a series of {\it triplets}, viz, 
$(\Phi[2], \Phi[2], \Phi[1])$.

(ii) This triplet repeats itself periodically as we sweep through the
points $\Phi_j=j/2^{m+1}\Phi_0$ along the flux axis between zero and half flux quantum. For 
any given value of $m$ which fixes the number of intervals, the cyclic invariance of the full 
parameter space will start showing up at a specific step of renormalization. 
This `step' $n$ is given by the `power' $m$ of $2$ in the denominator in the expression for the 
flux $\Phi$, whenever $\Phi=\frac{(2l+1)}{2^m}\Phi_0$, $l=0,1,2,...$. 

Thus, for an 
eight sub-interval splitting of the range $0 \le \Phi \le \Phi_0/2$, the first value (after zero) at 
$\Phi=\frac{1}{2^4}\Phi_0$ exhibits a $2$-cycle fixed point starting at $n=4$. The second flux 
$\Phi=\frac{2}{2^4}\Phi_0=\frac{1}{2^3}\Phi_0$ leads to a $2$-cycle fixed point beginning at 
$n=3$. For $\Phi=\frac{3}{2^4}\Phi_0$ we have a $1$-cycle fixed point for $n \ge 4$. This 
sequence of cycles repeats periodically, but the stage at which the invariance starts showing up
is not the same for every flux. However, for values of $m >2$, a subset of the fixed points 
($1$-cycle, or $2$-cycle) are truly equivalent in the sense that they start showing up at the 
same stage of RSRG. In this sense, the fixed points arising out of flux values 
$\Phi=\Phi_0/2^m$, $\Phi=5\Phi_0/2^m$, $\Phi=7\Phi_0/2^m$ are 
truly equivalent.

It should be appreciated that the sequence of flux values 
responsible for a certain cyclic behavior has a deterministic feature. If we 
select $\Phi=\frac{1}{2^m}\Phi_0$, with $m=2$, $3$, $4$,$...$ sequentially, then its easy to check that
for all such values of $\Phi$ we have a $2$-cycle fixed point beginning at $n=m$. Thus, at the 
left-most value of the flux, the cyclic invariance starts revealing at the deepest scale of length. As 
we move along the flux line, the fixed point character begins to show up earlier, and at $\Phi=\phi_0/4$, 
we see it immediately from $n=2$ onwards. Similarly, whenever $\Phi=\frac{3p}{2^m}\Phi_0$, with 
$p=1$, $2$, $3$,$...$, and $m \ge 2$, we come across a $1$-cycle fixed point for $n \ge m$.  
For example, by selecting $m=5$, the values of the flux at $\Phi/\Phi_0=3/{2^5}$, $6/2^5=3/2^4$, 
$9/2^5$, $12/2^5=3/2^3$, and $15/2^5$ exhibit $1$-cycle fixed point beginning at $n=5$, $4$, $5$, $3$ and 
$5$ respectively. Similar  deterministic feature has also been possible to locate for other sub-divisions.

It is obvious that the separation between the one cycle, and the two-cycle values of the 
flux becomes exponentially smaller as we split the flux interval between zero and one-half 
flux quantum more and more by increasing $m$. The behavior of ($\Phi[2], \Phi[2], \Phi[1])$
is consistent with our expectation, so far as we have tested. This encourages us to conclude 
that, speaking just in terms of the one and two-cycle fixed points, there will be a 
`quasi-continuous' cross-over in the character of the extended eigenstates as one sweeps over 
the specific flux values at $\Phi_j=\frac{j}{2^{m+1}}\Phi_0$ along the flux line 
between zero and half flux quantum. 

\begin{center}
{\it What happens in the range $\Phi_0/2 \le \Phi \le \Phi_0$ ?}
\end{center}

Similar correlations between the flux values and the 
cycles of the fixed points may also be looked for the flux ranging from $\Phi=\Phi_0/2$ to one 
flux quantum. The pattern however, may be different. 
For example, if we split the full range of flux, from $\Phi=0$ to $\Phi=\Phi_0$, in $2^5=32$ 
equal intervals, then it is seen that from $\Phi=j\Phi_0/32$, with $j=1,2,...,15$, the 
pattern $(\Phi[2], \Phi[2], \Phi[1])$ is periodically repeated. The mid-point $\Phi=\Phi_0$ 
has to be omitted as we do not get any state there. In the rest of the flux interval, 
that is, from $\Phi=17\Phi_0/32$ and upto $\Phi=31\Phi_0/32$ in interval of $\Phi_0/32$, we  
now get a triplet $(\Phi[2], \Phi[1], \Phi[2])$ repeating periodically. One can proceed in this 
way for other values of $m$. 
For any given $m$, the entire pattern observed between $\Phi=0$ and $\Phi=\Phi_0$ 
obviously repeats beyond one flux quantum. We have tested these observations, by working out 
the patterns for a few values of $m$ to begin with, and then speculated the pattern for
larger values of $m$. This test has been successful and gives us confidence to predict the
correlation as it has been described above. 

Before we end this section, it is good to note that, we have chosen to speak in terms of the 
one and the two cycle fixed points only. There are, other flux values as well for which one
gets different multiple cyclic behavior, even a completely chaotic behavior of the parameter space.
All these point towards the existence of extended eigenstates in a $3$-simplex network. However, 
we have tried to focus on a definite correlation between cycles of invariance of the parameter space, 
and a given set of values of the magnetic flux by citing the above example.
\section{Conclusion}
We have examined the spectral properties of a $3$-simplex fractal network in the 
presence of a magnetic field penetrating a subspace of this fractal space. Both the 
isotropic and the anisotropic limits of the model have been discussed with special 
emphasis on the flux dependent electronic transmission and the existence of extended electronic 
states. Based on a numerical study of the exact renormalization group recursion relations 
we show that there is a subtle correlation between the value of the magnetic flux and the 
fixed point behavior of the Hamiltonian, and propose that such an observation may lead to a 
method of classification of the extended eigenstates for a given value of the energy of the 
electron.
\begin{center}
{\bf Acknowledgment}
\end{center}
The author is grateful to the Max Planck Institute f\"{u}r Physik Complexer Systeme, in Dresden, 
for their kind hospitality and for supporting the present work. Special thanks must be 
given to Magnus Johansson for a critical reading of the manuscript and for valuable comments.
   
\end{document}